\documentclass[english,aps,superscriptaddress,preprint]{revtex4}
\usepackage[T1]{fontenc}
\usepackage{color}
\usepackage{array}
\usepackage{multirow}
\usepackage{amsmath}
\usepackage{amssymb}
\usepackage{graphicx}

\makeatletter

\providecommand{\tabularnewline}{\\}

\@ifundefined{textcolor}{}
{%
 \definecolor{BLACK}{gray}{0}
 \definecolor{WHITE}{gray}{1}
 \definecolor{RED}{rgb}{1,0,0}
 \definecolor{GREEN}{rgb}{0,1,0}
 \definecolor{BLUE}{rgb}{0,0,1}
 \definecolor{CYAN}{cmyk}{1,0,0,0}
 \definecolor{MAGENTA}{cmyk}{0,1,0,0}
 \definecolor{YELLOW}{cmyk}{0,0,1,0}
}

\usepackage{dcolumn}% needed for some tables
\usepackage{bm}% for math
\usepackage{times}
\usepackage{hyperref}

\makeatother

\usepackage{babel}
\begin{document}
\title{Ion parallel closures\\ \href{https://dx.doi.org/10.1063/1.4977054}{\small Journal-ref:  \underline{Phys.  Plasmas 24, 022127 (2017)}} {\small \color{red}  with corrections}}
\author{Jeong-Young Ji}
\email{j.ji@usu.edu}

\affiliation{Department of Physics, Utah State University, Logan, Utah 84322}
\author{Hankyu Q. Lee}
\affiliation{Department of Physics, Utah State University, Logan, Utah 84322}
\author{Eric D. Held}
\affiliation{Department of Physics, Utah State University, Logan, Utah 84322}
\date{\today}
\begin{abstract}
Ion parallel closures are obtained for arbitrary atomic weights and
charge numbers. For arbitrary collisionality, the heat flow and viscosity
are expressed as kernel-weighted integrals of the temperature and
flow-velocity gradients. Simple, fitted kernel functions are obtained
from the 1600 parallel moment solution and the asymptotic behavior
in the collisionless limit. The fitted kernel parameters are tabulated
for various temperature ratios of ions to electrons. The closures
can be used conveniently without solving the kinetic equation or higher
order moment equations in closing ion fluid equations.
\end{abstract}
\maketitle

\section{Introduction}

The ion fluid equations for density $(n)$, flow velocity $(\mathbf{V})$,
and temperature $(T)$ are closed by expressing heat flux density
(\textbf{$\mathbf{h}$}) and viscosity tensor $(\boldsymbol{\pi})$
in terms of fluid variables, $n$, $T$, and $\mathbf{V}$. The ion
friction force and collisional heating densities can be obtained from
those of electrons~\citep{Ji2013H}. For high collisionality, the
ion closures are formulated in Ref.~\citep{Braginskii1958,Braginskii1965}
with the ion-electron collision effects ignored. The results are generalized
and improved by including the ion-electron collision terms in Ref.~\citep{Ji2015H}.
For low collisionality the free streaming term plays an important
role and the parallel closures appear in integral form~\citep{Hammett1990P,Hazeltine1998,Held2001CHS,Held2003,Held2003CH,Held2004-5}.
With accurate collision terms adopted, the electron parallel closures
for arbitrary collisionality are obtained in Refs\@.~\citep{Ji2014H1,Ji2016KHN}.

The integral (non-local) closures enable fluid models to capture kinetic
effects in parallel transport. The closures are implemented in the
BOUT++~\citep{Dudson2009e4} to study kinetic effects on parallel
transport in fluid models of the scrape-off layer~\citep{Omotani2013D,Omotani2015DHU}.
The kernel functions obtained from the moment approach may be approximated
by a sum of modified-Helmholtz-equation solves in configuration space
for the fast non-Fourier method to compute closures efficiently~\citep{Dimits2014JU}. 

In this work we extend our previous work on parallel closures for
electrons~\citep{Ji2014H1,Ji2016KHN} to ions. We adopt the closure/transport
ordering ignoring the time derivative terms when solving general moment
equations for higher order moments. As the ion-electron collision
effects can be significant, we keep the ion-electron collision terms.
The ion-electron collision operator notably modifies closures for
high to moderate collisionality. The ion-electron collision terms
depend on the ion-electron temperature ratio, $T_{\mathrm{i}}/T_{\mathrm{e}}$,
and mass ratio combined with the ion charge number, $m_{\mathrm{e}}/m_{\mathrm{i}}Z^{2}=m_{\mathrm{e}}/m_{\mathrm{p}}AZ^{2}$,
where $m_{\mathrm{e}}$ is the electron mass, $m_{\mathrm{\text{i}}}$
is the ion mass, $m_{\mathrm{p}}$ is the proton mass, $A$ is the
atomic weight, and $Z$ is the ion charge number. We solve the general
moment equations for various temperature and mass ratios to obtain
kernels to compute closures. Then we construct simple fitted kernels
for arbitrary $AZ^{2}$ and temperature ratio $T_{\mathrm{i}}/T_{\mathrm{e}}\le10$.
For $AZ^{2}=1$ and 2, the fitted kernels are specified by seven parameters,
yielding highly accurate closures within 2\% errors. For $AZ^{2}\ge3$,
simpler form of kernels are specified by only four parameters which
are expressed as general functions of $AZ^{2}$ and $T_{\mathrm{i}}/T_{\mathrm{e}}$,
yielding accurate closures within 20\% errors.

In Sec.~\ref{sec:MEll}, we review the parallel moment equations
and the properties of kernels for the integral closures. In Sec.~\ref{sec:pll},
the fitted kernels and their parameters are presented for arbitrary
$AZ^{2}$. In Sec.~\ref{sec:Dis}, we summarize and discuss future
work.

\section{parallel moment equations and integral closures\label{sec:MEll}}

In this section we write a set of linearized parallel moment equations,
the solution of which provides closures. The derivations are basically
the same as the electron case in Sec. II of Ref.~\citep{Ji2016KHN}.
The parallel moment equations are obtained by taking parallel components
of the general moment equations~\citep{Ji2008H} or taking moments
of the following reduced drift kinetic equation
\begin{equation}
v_{\|}\frac{\partial\bar{f}_{\mathrm{i}}^{\mathrm{N}}}{\partial\ell}=C_{\mathrm{iL}}(\bar{f}_{\mathrm{i}}^{\mathrm{N}})-v_{\|}\frac{\partial\bar{f}_{\mathrm{i}}^{\mathrm{M}}}{\partial\ell}+C_{\mathrm{iL}}(\bar{f}_{\mathrm{i}}^{\mathrm{M}})\label{rKE}
\end{equation}
where $\ell$ is the arc length along the magnetic field line, $\partial/\partial\ell=\mathbf{b}\cdot\nabla$,
$\mathbf{b}=\mathbf{B}/B$, $\mathbf{B}$ is the magnetic field, and
$\bar{f}^{\mathrm{M}}$ and $\bar{f}^{\mathrm{N}}$ are the gyro-averaged
Maxwellian (M) and non-Maxwellian (N) distribution functions, respectively. 

The linearized collision operators $C_{\mathrm{iL}}$ are linearized
with respect to
\[
f_{a}^{\mathrm{m}}=\frac{n_{a}}{\pi^{3/2}v_{Ta}^{3}}e^{-s_{a}^{2}},
\]
where $n_{a}$ is the density of species $a$, $v_{Ta}=\sqrt{2T_{a}/m_{a}}$,
$s_{a}=v/v_{Ta}$, and $T_{a}$ is the temperature. For the non-Maxwellian
distribution,
\begin{equation}
C_{\mathrm{iL}}(\bar{f}_{\mathrm{i}}^{\mathrm{N}})=C(\bar{f}_{\mathrm{i}}^{\mathrm{N}},f_{\mathrm{i}}^{\mathrm{m}})+C(f_{\mathrm{i}}^{\mathrm{m}},\bar{f}_{\mathrm{i}}^{\mathrm{N}})+C(\bar{f}_{\mathrm{i}}^{\mathrm{N}},f_{\mathrm{e}}^{\mathrm{m}})\label{CiLN}
\end{equation}
 and for the Maxwellian
\begin{equation}
C_{\mathrm{iL}}(\bar{f}_{\mathrm{i}}^{\mathrm{M}})\approx C(f_{\mathrm{i}}^{\mathrm{m}},f_{\mathrm{e}}^{\mathrm{m}})+C(\bar{f}_{\mathrm{i}}^{\mathrm{M}-\mathrm{m}},f_{\mathrm{e}}^{\mathrm{m}})+C(f_{\mathrm{i}}^{\mathrm{m}},\bar{f}_{\mathrm{e}}^{\mathrm{M}-\mathrm{m}}),\label{CiLM}
\end{equation}
where
\[
\bar{f}_{a}^{\mathrm{M-m}}\approx2s_{a\|}\frac{V_{a\|}}{v_{Ta}}f_{a}^{\mathrm{m}}
\]
and $V_{a\|}$ is the parallel flow velocity. The ion-electron collision
operator for the Maxwellian distribution, Eq.~\eqref{CiLM}, which
equilibrates temperature and flow velocity between electrons and ions,
does not appear in the closure moment equations. Note that the ion-electron
collision effect is not ignorable~\citep{Ji2015H}. As explicitly
shown in Eqs.~(7), (9), and (12) of Ref.~\citep{Ji2015H}, the ion-electron
collision operator depends on $\sqrt{m_{\mathrm{e}}/m_{\mathrm{i}}}/Z=\sqrt{m_{\mathrm{e}}/m_{\mathrm{p}}AZ^{2}}$
and the temperature ratio $T_{\mathrm{e}}/T_{\mathrm{i}}$. 

The linearized parallel moment equations for the non-Maxwellian moments
are~\citep{Ji2016KHN} 
\begin{equation}
\sum_{lk\ne\mathrm{M}}\psi^{jp,lk}\frac{\partial n^{lk}}{\partial\eta}=\sum_{lk\ne\mathrm{M}}c^{jp,lk}n^{lk}+g^{jp},\label{ME}
\end{equation}
or, in matrix form,
\begin{equation}
\Psi\frac{d}{d\eta}[n]=C[n]+[g],\label{MEm}
\end{equation}
where $d\eta=d\ell/\lambda_{\mathrm{C}}$, $\lambda_{\mathrm{C}}=v_{T\mathrm{i}}\tau_{\mathrm{ii}}$,
and $\tau_{\mathrm{ii}}$ is the ion-ion collision time. Here and
hereafter the ion species index will be suppressed unless it is needed
for clarity. The matrix elements are
\begin{eqnarray}
\psi^{jp,lk} & = & \delta_{j+1,l}\psi_{pk}^{j}+\delta_{j-1,l}\psi_{kp}^{j-1},\nonumber \\
\psi_{pk}^{j} & = & \frac{j+1}{\sqrt{(2j+1)(2j+3)}}(\sqrt{j+p+\tfrac{3}{2}}\delta_{p,k}-\sqrt{p}\delta_{p-1,k}),\label{psi}
\end{eqnarray}
and
\[
c^{jp,lk}=\delta_{jl}\tau_{\mathrm{ii}}(\hat{A}_{\mathrm{ii}}^{jpk}+\hat{B}_{\mathrm{ii}}^{jpk}+\hat{A}_{\mathrm{ie}}^{jpk}),
\]
where $\hat{A}_{ab}^{jpk}$ and $\hat{B}_{ab}^{jpk}$ are explicitly
formulated in Ref.~\citep{Ji2008H}. The moment indices $(l,k)$
run 
\begin{eqnarray*}
(0,2),\;(0,3), & \cdots, & (0,K+1);\\
(1,1),\;(1,2), & \cdots, & (1,K);\\
(2,0),\;(2,1), & \cdots, & (2,K-1);\\
\vdots & \vdots & \vdots\\
(L,0),\;(L,1), & \cdots, & (L,K-1),
\end{eqnarray*}
excluding the Maxwellian moments $\mathrm{M}=(0,0),\ (0,1),$ and
$(1,0)$. 

The parallel closures are related to the general moments by 
\begin{eqnarray}
h_{\|} & = & -\frac{\sqrt{5}}{2}v_{T}Tn^{11},\label{h:n}\\
\pi_{\|} & = & \frac{2}{\sqrt{3}}Tn^{20}.\label{p:n}
\end{eqnarray}
For ions, the only non-vanishing thermodynamic drives are
\begin{eqnarray}
g_{\mathrm{i}}^{11} & = & \frac{\sqrt{5}}{2}\frac{n}{T}\frac{dT}{d\eta},\label{g1k}\\
g_{\mathrm{i}}^{20} & = & -\frac{\sqrt{3}}{2}n\tau_{\mathrm{ii}}W_{\|},\label{g20}
\end{eqnarray}
where
\begin{equation}
W_{\|}=\mathbf{b}\mathbf{b}:\mathsf{W},\;(\mathsf{W})_{\alpha\beta}=\partial_{\alpha}V_{\beta}+\partial_{\beta}V_{\alpha}-\frac{2}{3}\delta_{\alpha\beta}\nabla\cdot\mathbf{V}
\end{equation}
and $\mathbf{V}$ is the ion flow velocity. 

The system of $N$ moment equations can be solved by computing the
eigensystem of $\Psi^{-1}C$ where the eigenvalues appear in positive
and negative pairs. Different from the electron case, since the ion-electron
collision matrix is not symmetric, some eigenvalues are complex numbers.
The complex eigenvalues appear in complex conjugate pairs and so do
the corresponding eigenvectors, making the solution real. The solution
is expressed as a kernel weighted integral of the thermodynamic drives
\begin{equation}
n_{A}(z)=\sum_{D}\int_{-\infty}^{\infty}K_{AD}(z-z^{\prime})g_{D}(z^{\prime})dz^{\prime},\label{n:K}
\end{equation}
where the moment indices have been abbreviated as a single index $A,\;B,$
etc. The kernel functions are 
\begin{equation}
K_{AD}(\eta)=\begin{cases}
{\displaystyle -\sum_{\{B|\mathfrak{R}(k_{B})>0\}}^{N}\gamma_{AD}^{B}e^{k_{B}\eta}}, & \eta<0,\\
{\displaystyle +\sum_{\{B|\mathfrak{R}(k_{B})<0\}}^{N}\gamma_{AD}^{B}e^{k_{B}\eta}}, & \eta>0,
\end{cases}\label{K:e}
\end{equation}
where $\mathfrak{R}(k_{B})$ denotes the real part of the eigenvalue
$k_{B}$. The coefficients are 
\begin{equation}
\gamma_{AD}^{B}=\sum_{C}W_{AB}W_{BC}^{-1}\Psi_{CD}^{-1},\label{a:W}
\end{equation}
where $W_{AB}$ is the $A$-th component of the eigenvector with eigenvalue
$k_{B}$. 

For closure moments, we define 
\begin{eqnarray}
\gamma_{hh}^{B} & = & \frac{5}{2}\gamma_{11,11}^{B},\nonumber \\
\gamma_{h\pi}^{B} & = & -\sqrt{\frac{5}{3}}\gamma_{11,20}^{B},\nonumber \\
\gamma_{\pi h}^{B} & = & -\sqrt{\frac{5}{3}}\gamma_{20,11}^{B},\nonumber \\
\gamma_{\pi\pi}^{B} & = & \frac{4}{3}\gamma_{20,20}^{B},\label{gaAB}
\end{eqnarray}
and corresponding $K_{AD}$ by Eq.~\eqref{K:e}. The sign should
be corrected for $\gamma_{h\pi}^{B}$ and $\gamma_{R\pi}^{B}$ in
Eq. (35) of Ref. 12. Noting that
\begin{equation}
\gamma_{AD}^{-B}=\begin{cases}
-\gamma_{AD}^{B}, & AD=hh,\pi\pi\equiv\mathrm{even},\\
+\gamma_{AD}^{B}, & AD=h\pi,\pi h\equiv\mathrm{odd},
\end{cases}\label{gam+-}
\end{equation}
where $-B$ denotes the moment index corresponding to $-k_{B}$, we
notice that the kernel functions are even or odd functions
\begin{equation}
K_{AD}(-\eta)=\begin{cases}
+K_{AD}(\eta), & AD=\mathrm{even}\\
-K_{AD}(\eta), & AD=\mathrm{odd}.
\end{cases}\label{K+-}
\end{equation}
Using the definition of $K_{AD}$ and Eqs.~(\ref{h:n}-\ref{g20}),
we can write the parallel closures 
\begin{eqnarray}
h_{\|}(\ell) & = & Tv_{T}\int d\eta^{\prime}\Bigl(-\frac{1}{2}K_{hh}\frac{n}{T}\frac{dT}{d\eta^{\prime}}-K_{h\pi}\frac{3}{4}n\tau W_{\|}\Bigr),\label{h:}\\
\pi_{\|}(\ell) & = & T\int d\eta^{\prime}\Bigl(-K_{\pi h}\frac{n}{T}\frac{dT}{d\eta^{\prime}}-K_{\pi\pi}\frac{3}{4}n\tau W_{\|}\Bigr).\label{p:}
\end{eqnarray}

For sinusoidal drives, $T=T_{0}+T_{1}\sin\varphi$ and $V_{\|}=V_{0}+V_{1}\sin\varphi$,
where $\varphi=2\pi\ell/\lambda+\varphi_{0}=k\eta+\varphi_{0}$ and
$k=2\pi\lambda_{\mathrm{C}}/\lambda$, and assuming that $n$ and
$v_{T}\approx\sqrt{2T_{0}/m}$ are constant and $\nabla\cdot\mathbf{V}_{\perp}=0$,
the linearized closures become 
\begin{eqnarray}
h_{\|}(\ell) & = & -\frac{1}{2}nT_{1}v_{T}\hat{h}_{h}\cos\varphi-nT_{0}V_{1}\hat{h}_{\pi}\sin\varphi,\\
\pi_{\|}(\ell) & = & -nT_{1}\hat{\pi}_{h}\sin\varphi-nT_{0}\frac{V_{1}}{v_{T}}\hat{\pi}_{\pi}\cos\varphi.
\end{eqnarray}
The dimensionless closures are defined by $\hat{h}_{h}=k\hat{K}_{hh},$
$\hat{h}_{\pi}=k\hat{K}_{h\pi},$ $\hat{\pi}_{h}=k\hat{K}_{\pi h}$,
and $\hat{\pi}_{\pi}=k\hat{K}_{\pi\pi}$, where 
\begin{equation}
\hat{K}_{AD}=\begin{cases}
-\frac{1}{2}{\displaystyle \sum_{B=1}^{N}\gamma_{AD}^{B}\left(\frac{1}{k_{B}+ik}+\frac{1}{k_{B}-ik}\right)}, & AD=\mbox{even}\\
{\displaystyle \frac{i}{2}\sum_{B=1}^{N}\gamma_{AD}^{B}\left(\frac{1}{k_{B}+ik}-\frac{1}{k_{B}-ik}\right)}, & AD=\mbox{odd},
\end{cases}\label{Khad}
\end{equation}
which are derived from Eqs.~\eqref{K:e}, \eqref{gam+-}, and 
\begin{equation}
\int K_{AD}(\eta-\eta^{\prime})\cos(k\eta^{\prime}+\varphi_{0})d\eta^{\prime}=\begin{cases}
\hat{K}_{AD}\cos\varphi, & AD=\mbox{even},\\
{\displaystyle \hat{K}}_{AD}\sin\varphi, & AD=\mbox{odd}.
\end{cases}
\end{equation}

\begin{figure}
\includegraphics{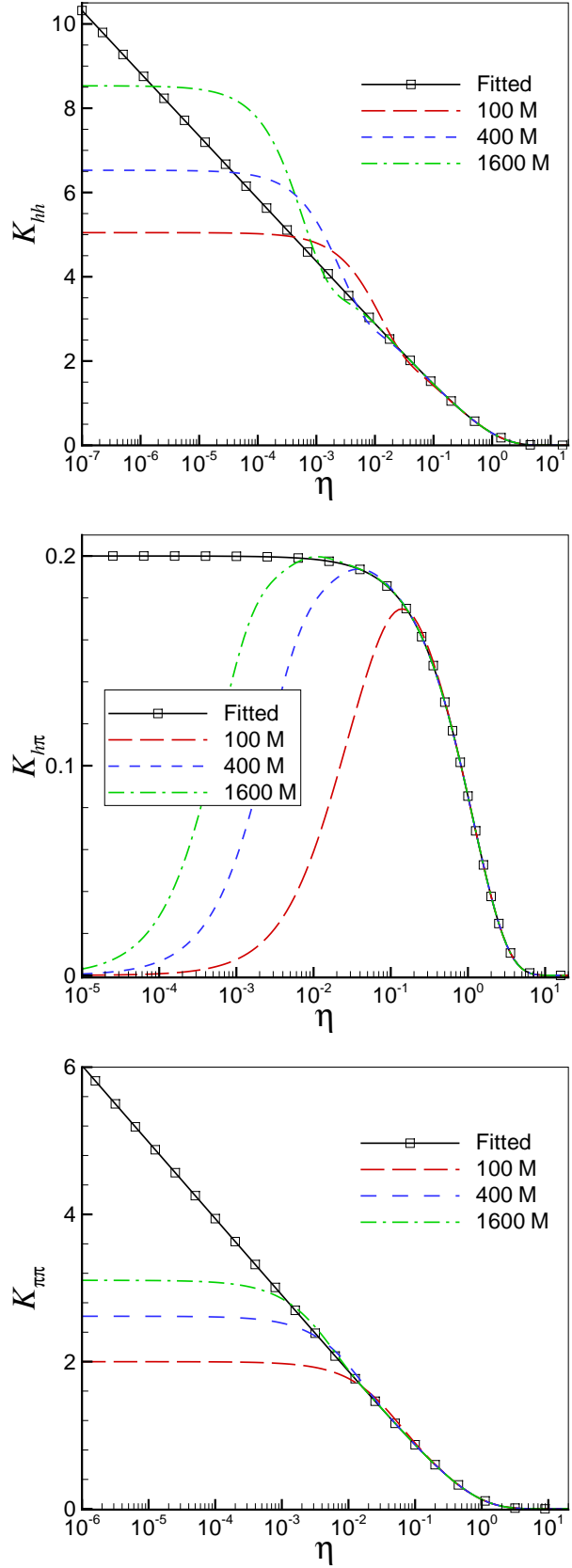}\caption{Kernels for $AZ^{2}=1$ and $T_{\mathrm{i}}/T_{\mathrm{e}}=4$. The
kernel $K_{\pi h}$ (not shown) is similar to $K_{h\pi}.$ }
\label{fig1}
\end{figure}
 Fig.~\ref{fig1} shows typical behavior of the kernels from $N=100\,(L=10,\,K=10)$,
$N=400\,(L=20,\,K=20)$, and $N=1600\ (L=40,\,K=40)$ moment calculations.
\begin{figure}
\includegraphics{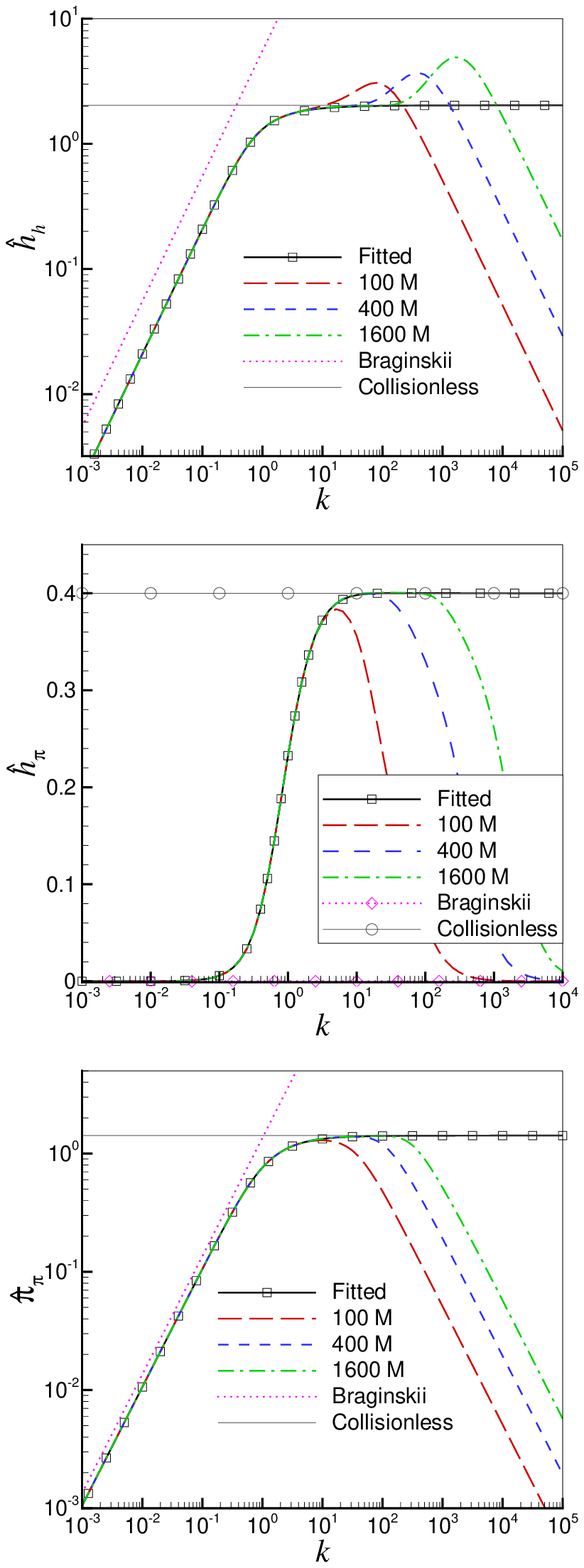}\caption{Closures for $AZ^{2}=1$ and $T_{\mathrm{i}}/T_{\mathrm{e}}=4$. The
closure $\hat{\pi}_{h}$ (not shown) is similar to $\hat{h}_{\pi}$.}
\label{fig2}
\end{figure}
 With increasing number of moments, the kernels converge for smaller
$\eta$. However, convergence is slow. For $\eta\ll1$, analytical
calculations show that the kernels approach~\citep{Ji2013HJ}
\begin{eqnarray}
K_{hh}(\eta) & \approx & -\frac{18}{5\pi^{3/2}}(\ln|\eta|+\gamma_{h}),\label{cless11}\\
K_{h\pi}(\eta) & \approx & \frac{1}{5},\\
K_{\pi h}(\eta) & \approx & \frac{1}{5},\\
K_{\pi\pi}(\eta) & \approx & -\frac{4}{5\pi^{1/2}}(\ln|\eta|+\gamma_{\pi}),\label{cless22}
\end{eqnarray}
where $\gamma_{h}$ and $\gamma_{\pi}$ are constants. Fig.~\ref{fig2}
shows the corresponding closures. The $N=1600$ moment closures converge
for $k\lesssim100$ and reach collisionless values at $k\sim30$.
This situation is very different from the electron case where even
the $N=6400$ moment closures do not reach the collisionless values
in the convergent regime $k\lesssim100$ (see Fig.~2 of Ref.~\citep{Ji2014H1}).
The convergence of the electron closures is slower than the ion closures.
This is because the coupling between moments of different velocity
orders in the electron-ion collision operator is much larger than
that in the ion-electron operator. Deviations of moment closures for
large $k\gtrsim100$ will be amended by collisionless kernels for
small $\eta$. Finally the corrections to Braginskii provided by the
moment closures, $\hat{h}_{h}$ and $\hat{\pi}_{\pi}$, even in the
highly collisional regime are due to the inclusion of the ion-electron
collision operator. 

\section{Fitted kernels for integral closures\label{sec:pll}}

As shown in Figs.~\ref{fig1} and \ref{fig2}, the kernels obtained
from a finite number of moment equations are not accurate for small
$\eta$ and result in inaccurate closures in the collisionless limit.
Furthermore they involve as many terms as the number of moments used
in the derivation {[}see Eq.~\eqref{K:e}{]}. Therefore simple fitted
functions are desirable to accurately represent the moment kernels
for the convergent regime and the collisionless kernels for small
$\eta$. Due to the ion-electron collision operator in Eq.~\eqref{CiLN},
the kernels depend on $AZ^{2}$ and $T_{\mathrm{i}}/T_{\mathrm{e}}$.
By computing the kernels and closures for $AZ^{2}=1,2,3,4,16,\cdots,4096$
and $T_{\mathrm{i}}/T_{\mathrm{e}}=0,\;0.01,\;0,1,\;1,\ 2,\cdots,\;10$,
highly accurate fitted kernels for arbitrary $AZ^{2}$ and $T_{\mathrm{i}}/T_{\mathrm{e}}\lesssim10$
may be obtained via interpolation. 

\subsection{Kernels for $AZ^{2}=1\,(\mathrm{H}^{+})$ and $2\,(\mathrm{D}^{+})$}

As in the electron case, all kernel functions can be fitted to a single
function
\begin{equation}
K_{AB}(\eta)=-[d+a\exp(-b\eta^{c})]\ln[1-\alpha\exp(-\beta\eta^{\gamma})]\label{Kfit}
\end{equation}
which yields highly accurate closures for arbitrary collisionality.
In order to reproduce the collisionless-limit kernels (\ref{cless11}-\ref{cless22}),
the parameter $a$ is derived from other parameters 
\begin{eqnarray}
a & = & \frac{18}{5\pi^{3/2}\gamma}-d\mbox{ for }K_{hh},\label{ahh}\\
a & = & \frac{4}{5\pi^{1/2}\gamma}-d\mbox{ for }K_{\pi\pi},\label{app}
\end{eqnarray}
and
\begin{equation}
a=-d-\frac{1}{5\ln(1-\alpha)}\mbox{ for }K_{h\pi}\text{ and }K_{h\pi}.\label{ahp}
\end{equation}
In the collisional limit, the kernels also reproduce Braginskii-type
parallel closures~\citep{Ji2015H} 
\begin{eqnarray}
h_{\|} & = & -\hat{\kappa}_{\|}\frac{nT\tau_{\mathrm{ii}}}{m}\partial_{\|}T,\label{hc:}\\
\pi_{\|} & = & -\hat{\eta}_{0}nT\tau_{\mathrm{ii}}W_{\|}\label{pc:}
\end{eqnarray}
with improved coefficients by including ion-electron collision effects
and more moments. 

In the $AZ^{2}\rightarrow\infty$ and/or $T_{\mathrm{i}}/T_{\mathrm{e}}\rightarrow0$
limits, the ion-electron collision terms vanish and the fitted parameters
are presented in Table \ref{t:K}. In the collisional limit, the closure
coefficients become Braginskii's coefficients $\hat{\kappa}_{\|}=\int_{-\infty}^{\infty}K_{hh}d\eta\approx5.586$
and $\hat{\eta}_{0}=\int_{-\infty}^{\infty}\frac{3}{4}K_{\pi\pi}d\eta\approx1.365$
where the values are slightly improved from Braginskii's due to the
increased number of moments. 

\begin{table}
\begin{tabular}{ccccccccc}
\hline 
 & $a$ & $b$ & c & $d$ & $\alpha$ & $\beta$ & $\gamma$ & err.\tabularnewline
\hline 
$K_{hh}$ & 0.141 & 1.86 & 0.721 & 0.974 & 1 & 0.823 & 0.58 & 0.2\%\tabularnewline
\textcolor{red}{$K_{\pi h}$} & -0.750 & 1.23 & 0.600 & 1.10 & 0.434 & 1.28 & 0.502 & 1.6\%\tabularnewline
\textcolor{red}{$K_{h\pi}$} & -0.701 & 1.30 & 0.602 & 1.07 & 0.42 & 1.24 & 0.510 & 0.6\%\tabularnewline
$K_{\pi\pi}$ & 0.440 & 0.641 & 0.791 & 0.319 & 1 & 1.09 & 0.595 & 0.3\%\tabularnewline
\hline 
\end{tabular}\caption{Fitted parameters in Eq.~\eqref{Kfit} with no ion-electron collision
operator (in the $AZ^{2}\rightarrow\infty$ and/or $T_{\mathrm{i}}/T_{\mathrm{e}}\rightarrow0$
limits).}
\label{t:K}
\end{table}

As $T_{\mathrm{i}}/T_{\mathrm{e}}$ increases, ion-electron collisions
become significant. The effect is more significant for smaller values
of $AZ^{2}$ due to the factor $1/\sqrt{AZ^{2}}$ in the ion-electron
collision terms. The fitted parameters for $AZ^{2}=1\,(\mathrm{H}^{+})$
and $2\,(\mathrm{D}^{+})$ are presented in Tables \ref{t:K01} and
\ref{t:K02}, respectively. The fitted kernels for $AZ^{2}=1$ and
$T_{\mathrm{i}}/T_{\mathrm{e}}=4$ are shown in Fig.~\ref{fig1}
and the corresponding closures are shown in Fig.~\ref{fig2}. Note
that computed closures approach collisional and collisionless closures
in the $k\rightarrow0$ and $\infty$ limits, respectively. 

To evaluate the accuracy of fitted kernels, closures computed from
the fitted kernels are compared with 1600 moment closures computed
from Eq.~\eqref{Khad} in the convergent regime $k\lesssim80$. Maximum
percentage errors are at most 1.9\% at a specific temperature ratio
and less than 1\% at most temperature ratios as shown in Tables \ref{t:K01}
and \ref{t:K02}. For a temperature ratio $t=T_{\mathrm{i}}/T_{\mathrm{e}}$,
not listed in the Tables \ref{t:K01} and \ref{t:K02}, parameters
can be obtained from a simple linear interpolation between two temperature
ratios $t_{1}$ and $t_{2}$ ($t_{1}<t<t_{2}$)
\begin{equation}
\Gamma(t)=\frac{t_{2}-t}{t_{2}-t_{1}}\Gamma(t_{1})+\frac{t-t_{1}}{t_{2}-t_{1}}\Gamma(t_{2})\label{At}
\end{equation}
for $\Gamma=b,c,d,\alpha,\beta,\gamma$. The parameter $a$ can be
obtained from the interpolated values by using Eqs.~\eqref{ahh},
\eqref{app} and \eqref{ahp}. The closures computed from the interpolated
parameters show similar accuracy. For $T_{\mathrm{i}}/T_{\mathrm{e}}<0.01$,
the ion-electron collision effect is ignorable and the parameters
of Table \ref{t:K} produce accurate closures within 2\% error. \renewcommand\arraystretch{0.7}
\begin{table}
\begin{tabular}{cccccccccccccc}
\hline 
$K_{AB}$ & $T_{\mathrm{i}}/T_{\mathrm{e}}$ & $0.01$ & $0.1$ & 1 & 2 & 3 & 4 & 5 & 6 & 7 & 8 & 9 & 10\tabularnewline
\hline 
\multirow{8}{*}{$K_{hh}$} & $a$ & 0.161 & 0.190 & 0.130 & 0.0723 & 0.0353 & 0.0116 & -0.00597 & -0.0179 & -0.0209 & 0.000474 & 0.0145 & 0.0234\tabularnewline
 & $b$ & 1.77 & 1.77 & 1.77 & 1.79 & 1.81 & 1.82 & 1.82 & 1.82 & 1.82 & 1.74 & 1.59 & 1.51\tabularnewline
 & $c$ & 0.660 & 0.646 & 0.630 & 0.610 & 0.581 & 0.554 & 0.537 & 0.519 & 0.5 & 0.461 & 0.416 & 0.363\tabularnewline
 & $d$ & 0.954 & 0.942 & 0.930 & 0.850 & 0.780 & 0.719 & 0.669 & 0.625 & 0.579 & 0.500 & 0.435 & 0.375\tabularnewline
 & $\alpha$ & 1 & 1 & 1 & 1 & 1 & 1 & 1 & 1 & 1 & 1 & 1 & 1\tabularnewline
 & $\beta$ & 0.811 & 0.814 & 0.857 & 0.907 & 0.993 & 1.11 & 1.27 & 1.47 & 1.71 & 1.96 & 2.30 & 2.81\tabularnewline
 & $\gamma$ & 0.579 & 0.570 & 0.61 & 0.701 & 0.793 & 0.885 & 0.975 & 1.07 & 1.16 & 1.29 & 1.44 & 1.62\tabularnewline
 & err. & 0.6\% & 1.0\% & 0.5\% & 0.7\% & 0.6\% & 0.6\% & 0.5\% & 0.7\% & 0.5\% & 0.6\% & 0.6\% & 0.6\%\tabularnewline
\hline 
\multirow{8}{*}{\textcolor{red}{$K_{\pi h}$}} & $a$ & -0.750 & -0.678 & -0.405 & -0.146 & -0.0296 & 0.0295 & 0.0650 & 0.119 & 0.160 & 0.218 & 0.280 & 0.331\tabularnewline
 & $b$ & 1.23 & 1.45 & 1.45 & 1.56 & 1.68 & 1.78 & 1.94 & 1.95 & 1.96 & 1.96 & 1.95 & 1.96\tabularnewline
 & $c$ & 0.600 & 0.616 & 0.624 & 0.671 & 0.780 & 0.830 & 0.899 & 0.923 & 0.946 & 0.981 & 1.02 & 1.03\tabularnewline
 & $d$ & 1.10 & 1.04 & 0.766 & 0.521 & 0.448 & 0.419 & 0.418 & 0.398 & 0.356 & 0.315 & 0.252 & 0.243\tabularnewline
 & $\alpha$ & 0.435 & 0.426 & 0.425 & 0.413 & 0.380 & 0.36 & 0.339 & 0.321 & 0.321 & 0.313 & 0.313 & 0.295\tabularnewline
 & $\beta$ & 1.28 & 1.25 & 1.05 & 0.791 & 0.683 & 0.677 & 0.736 & 0.775 & 0.839 & 0.903 & 0.952 & 1.13\tabularnewline
 & $\gamma$ & 0.502 & 0.501 & 0.585 & 0.765 & 0.943 & 1.08 & 1.19 & 1.33 & 1.46 & 1.61 & 1.79 & 1.88\tabularnewline
 & err. & 1.5\% & 1.9\% & 1.0\% & 0.5\% & 0.6\% & 0.5\% & 0.6\% & 0.3\% & 0.3\% & 0.3\% & 0.3\% & 0.5\%\tabularnewline
\hline 
\multirow{8}{*}{\textcolor{red}{$K_{h\pi}$}} & $a$ & -1.06 & -1.19 & -0.789 & -0.354 & -0.0270 & 0.221 & 0.574 & 1.16 & 1.57 & 1.70 & 1.21 & 0.951\tabularnewline
 & $b$ & 1.37 & 1.41 & 1.58 & 1.72 & 1.93 & 2.03 & 2.11 & 2.12 & 2.13 & 2.70 & 3.10 & 3.53\tabularnewline
 & $c$ & 0.604 & 0.611 & 0.619 & 0.644 & 0.693 & 0.723 & 0.739 & 0.774 & 0.793 & 0.817 & 0.818 & 0.819\tabularnewline
 & $d$ & 1.73 & 1.87 & 1.72 & 1.47 & 1.18 & 1.06 & 0.829 & 0.338 & -0.0193 & -0.0321 & -0.0478 & -0.0543\tabularnewline
 & $\alpha$  & 0.257 & 0.254 & 0.194 & 0.164 & 0.159 & 0.145 & 0.133 & 0.125 & 0.121 & 0.113 & 0.158 & 0.2\tabularnewline
 & $\beta$ & 1.24 & 1.29 & 1.06 & 0.937 & 0.934 & 1.03 & 1.09 & 0.827 & 0.326 & 0.331 & 0.513 & 0.76\tabularnewline
 & $\gamma$  & 0.504 & 0.490 & 0.584 & 0.725 & 0.857 & 0.965 & 1.12 & 1.59 & 1.62 & 1.86 & 2.07 & 1.97\tabularnewline
 & err. & 1.4\% & 1.5\% & 0.3\% & 0.4\% & 0.3\% & 0.7\% & 0.5\% & 0.8\% & 1.3\% & 1.3\% & 1.0\% & 1.1\%\tabularnewline
\hline 
\multirow{8}{*}{$K_{\pi\pi}$} & $a$ & 0.448 & 0.439 & 0.392 & 0.248 & 0.133 & 0.0801 & 0.0398 & 0.0193 & $6.00\times10^{-5}$ & -0.0154 & -0.0359 & -0.0512\tabularnewline
 & $b$ & 0.642 & 0.657 & 0.657 & 0.738 & 0.966 & 1.11 & 1.42 & 1.48 & 1.55 & 1.66 & 1.70 & 1.81\tabularnewline
 & $c$ & 0.788 & 0.788 & 0.810 & 0.882 & 0.972 & 1.01 & 1.05 & 1.12 & 1.14 & 1.15 & 1.26 & 1.28\tabularnewline
 & $d$ & 0.316 & 0.324 & 0.340 & 0.407 & 0.461 & 0.469 & 0.474 & 0.468 & 0.463 & 0.458 & 0.459 & 0.458\tabularnewline
 & $\alpha$ & 1 & 1 & 1 & 1 & 1 & 1 & 1 & 1 & 1 & 1 & 1 & 1\tabularnewline
 & $\beta$ & 1.09 & 1.09 & 1.134 & 1.22 & 1.34 & 1.48 & 1.65 & 1.84 & 2.06 & 2.32 & 2.63 & 2.97\tabularnewline
 & $\gamma$ & 0.591 & 0.591 & 0.617 & 0.689 & 0.761 & 0.822 & 0.878 & 0.927 & 0.975 & 1.02 & 1.07 & 1.11\tabularnewline
 & err. & 0.5\% & 0.4\% & 0.6\% & 0.4\% & 0.4\% & 0.3\% & 0.4\% & 0.3\% & 0.3\% & 0.2\% & 0.3\% & 0.2\%\tabularnewline
\hline 
\end{tabular}\caption{Fitted parameters in Eq.~\eqref{Kfit} for $AZ^{2}=1$.}
\label{t:K01}
\end{table}
 \renewcommand\arraystretch{0.7}
\begin{table}
\begin{tabular}{cccccccccccccc}
\hline 
$K_{AB}$ & $T_{\mathrm{i}}/T_{\mathrm{e}}$ & $0.01$ & $0.1$ & 1 & 2 & 3 & 4 & 5 & 6 & 7 & 8 & 9 & 10\tabularnewline
\hline 
\multirow{8}{*}{$K_{hh}$} & $a$ & 0.144 & 0.159 & 0.113 & 0.0634 & 0.0316 & 0.00509 & -0.0217 & -0.0422 & -0.0540 & -0.0845 & -0.121 & -1.24\tabularnewline
 & $b$ & 1.94 & 2.20 & 2.42 & 2.7 & 2.91 & 1.72 & 1.23 & 0.9 & 0.843 & 0.597 & 0.434 & 0.393\tabularnewline
 & $c$ & 0.720 & 0.735 & 0.749 & 0.77 & 0.8 & 1.00 & 1.47 & 1.47 & 1.40 & 1.29 & 1.25 & 1.2\tabularnewline
 & $d$ & 0.973 & 0.969 & 0.955 & 0.898 & 0.837 & 0.788 & 0.736 & 0.691 & 0.650 & 0.633 & 0.630 & 0.6\tabularnewline
 & $\alpha$ & 1 & 1 & 1 & 1 & 1 & 1 & 1 & 1 & 1 & 1 & 1 & 1\tabularnewline
 & $\beta$ & 0.821 & 0.823 & 0.846 & 0.887 & 0.946 & 1.03 & 1.12 & 1.22 & 1.36 & 1.53 & 1.75 & 2.01\tabularnewline
 & $\gamma$ & 0.579 & 0.573 & 0.606 & 0.672 & 0.744 & 0.815 & 0.905 & 0.996 & 1.09 & 1.18 & 1.27 & 1.36\tabularnewline
 & err. & 0.2\% & 0.9\% & 0.4\% & 0.6\% & 0.6\% & 0.7\% & 0.4\% & 0.4\% & 0.3\% & 0.3\% & 0.3\% & 0.4\%\tabularnewline
\hline 
\multirow{8}{*}{\textcolor{red}{$K_{\pi h}$}} & $a$ & -0.750 & -0.688 & -0.431 & -0.169 & -0.0524 & -0.00959 & 0.0170 & 0.0454 & 0.0659 & 0.108 & 0.153 & 0.206\tabularnewline
 & $b$ & 1.23 & 1.40 & 1.43 & 1.62 & 1.7 & 1.83 & 1.84 & 1.86 & 1.85 & 1.85 & 1.85 & 1.86\tabularnewline
 & $c$ & 0.600 & 0.608 & 0.619 & 0.670 & 0.673 & 0.673 & 0.777 & 0.880 & 0.895 & 0.955 & 0.990 & 1.02\tabularnewline
 & $d$ & 1.10 & 1.07 & 0.789 & 0.514 & 0.367 & 0.303 & 0.274 & 0.243 & 0.222 & 0.208 & 0.209 & 0.204\tabularnewline
 & $\alpha$ & 0.434 & 0.409 & 0.428 & 0.44 & 0.47 & 0.494 & 0.497 & 0.500 & 0.500 & 0.470 & 0.424 & 0.386\tabularnewline
 & $\beta$ & 1.28 & 1.24 & 1.06 & 0.807 & 0.661 & 0.620 & 0.618 & 0.615 & 0.641 & 0.646 & 0.698 & 0.749\tabularnewline
 & $\gamma$ & 0.502 & 0.504 & 0.573 & 0.722 & 0.882 & 1.01 & 1.13 & 1.26 & 1.37 & 1.52 & 1.64 & 1.77\tabularnewline
 & err. & 1.5\% & 1.2\% & 1.5\% & 0.5\% & 0.4\% & 0.3\% & 0.2\% & 0.2\% & 0.6\% & 0.5\% & 0.3\% & 0.3\%\tabularnewline
\hline 
\multirow{8}{*}{\textcolor{red}{$K_{h\pi}$}} & $a$ & -1.06 & -1.13 & -0.577 & -0.273 & -0.119 & 0.0440 & 0.229 & 0.363 & 0.521 & 0.686 & 0.638 & 0.440\tabularnewline
 & $b$ & 1.37 & 1.42 & 1.69 & 1.84 & 1.84 & 1.84 & 1.84 & 1.97 & 2.18 & 2.40 & 2.41 & 2.71\tabularnewline
 & $c$ & 0.599 & 0.613 & 0.630 & 0.659 & 0.691 & 0.750 & 0.749 & 0.758 & 0.797 & 0.844 & 0.823 & 0.817\tabularnewline
 & $d$ & 1.75 & 1.82 & 1.27 & 0.975 & 0.821 & 0.658 & 0.473 & 0.338 & 0.182 & 0.0534 & -0.00799 & -0.0116\tabularnewline
 & $\alpha$  & 0.253 & 0.254 & 0.251 & 0.248 & 0.248 & 0.248 & 0.248 & 0.248 & 0.248 & 0.237 & 0.272 & 0.373\tabularnewline
 & $\beta$ & 1.23 & 1.27 & 1.03 & 0.900 & 0.895 & 0.892 & 0.853 & 0.853 & 0.704 & 0.35 & 0.209 & 0.276\tabularnewline
 & $\gamma$  & 0.509 & 0.496 & 0.580 & 0.699 & 0.799 & 0.915 & 1.07 & 1.24 & 1.60 & 2.61 & 2.18 & 2.26\tabularnewline
 & err. & 0.8\% & 1.7\% & 0.5\% & 0.7\% & 0.6\% & 0.2\% & 0.5\% & 0.4\% & 0.6\% & 0.8\% & 1.2\% & 0.6\%\tabularnewline
\hline 
\multirow{8}{*}{$K_{\pi\pi}$} & $a$ & 0.446 & 0.441 & 0.405 & 0.294 & 0.186 & 0.106 & 0.0558 & 0.0255 & 0.0139 & -0.0116 & -0.0284 & -0.0661\tabularnewline
 & $b$ & 0.641 & 0.652 & 0.652 & 0.692 & 0.816 & 1.07 & 1.54 & 2.64 & 1.22 & 1.10 & 1.07 & 1.04\tabularnewline
 & $c$ & 0.789 & 0.789 & 0.805 & 0.856 & 0.923 & 1.00 & 1.10 & 1.26 & 1.79 & 1.89 & 1.67 & 1.67\tabularnewline
 & $d$ & 0.317 & 0.321 & 0.334 & 0.384 & 0.436 & 0.471 & 0.486 & 0.488 & 0.478 & 0.480 & 0.478 & 0.495\tabularnewline
 & $\alpha$ & 1 & 1 & 1 & 1 & 1 & 1 & 1 & 1 & 1 & 1 & 1 & 1\tabularnewline
 & $\beta$ & 1.09 & 1.09 & 1.12 & 1.19 & 1.27 & 1.37 & 1.48 & 1.62 & 1.77 & 1.94 & 2.12 & 2.35\tabularnewline
 & $\gamma$ & 0.592 & 0.592 & 0.611 & 0.666 & 0.725 & 0.782 & 0.833 & 0.879 & 0.918 & 0.965 & 1.00 & 1.05\tabularnewline
 & err. & 0.6\% & 0.3\% & 0.3\% & 0.9\% & 0.5\% & 0.4\% & 0.4\% & 0.4\% & 0.7\% & 0.4\% & 0.3\% & 0.3\%\tabularnewline
\hline 
\end{tabular}\caption{Fitted parameters in Eq.~\eqref{Kfit} for $AZ^{2}=2$.}
\label{t:K02}
\end{table}

\subsection{Kernels for $AZ^{2}\ge3$}

One can continue with Eq.~\eqref{Kfit} to obtain fitted parameters
which are accurate to within 2\% error for arbitrary $AZ^{2}\ge3$.
When $AZ^{2}$ is large, however, the ion-electron collision effect
becomes less significant and a simpler fitted function may suffice
to represent the moment kernels with collisionless asymptotes. For
$AZ^{2}\ge3$, we adopt the following form for the fitted kernels
\begin{equation}
K_{AB}(\eta)=-\kappa\ln[1-\alpha\exp(-\beta\eta^{\gamma})].\label{Kred}
\end{equation}
Using the parameters $\alpha$, $\beta$, and $\gamma$ obtained for
$AZ^{2}=3,\;16,\;64$ and for $T_{\mathrm{i}}/T_{\mathrm{e}}=0,\,1,\,5,\,9$,
the following interpolation formula is obtained for $\Gamma=\alpha,\beta,\gamma$
\begin{equation}
\Gamma(t,x)=(a_{33}t^{2}+a_{32}t+a_{31})tx^{3}+(a_{23}t^{2}+a_{22}t+a_{21})tx^{2}+(a_{13}t^{2}+a_{12}t+a_{11})tx+a_{00}\label{Ga}
\end{equation}
where $t=T_{\mathrm{i}}/T_{\mathrm{e}}$ and $x=1/\sqrt{AZ^{2}}$.
The coefficients $a_{ij}$ for $AB=hh,\,h\pi,\,\pi h,$ and $\pi\pi$
are presented in Table \ref{t:Kred}. The parameter $\kappa$ can
be obtained from the collisionless constraints (\ref{cless11}-\ref{cless22}),
\begin{equation}
\kappa=\begin{cases}
{\displaystyle \frac{18}{5\pi^{3/2}\gamma}}, & \mbox{ for }K_{hh}\\
{\displaystyle \frac{4}{5\pi^{1/2}\gamma}}, & \mbox{ for }K_{\pi\pi}\\
{\displaystyle -\frac{1}{5\ln(1-\alpha)},} & \mbox{ for }K_{h\pi}\text{ and }K_{\pi h}
\end{cases}.\label{clessk}
\end{equation}
 
\begin{table}
\begin{tabular}{c|c|cccccccccc}
\hline 
$K_{AB}$ & $\Gamma$ & $a_{00}$ & $a_{11}$ & $a_{12}$ & $a_{13}$ & $a_{21}$ & $a_{22}$ & $a_{23}$ & $a_{31}$ & $a_{32}$ & $a_{33}$\tabularnewline
\hline 
\multirow{3}{*}{$K_{hh}$} & $\alpha$ & 1 & 0 & 0 & 0 & 0 & 0 & 0 & 0 & 0 & 0\tabularnewline
 & $\beta$ & 0.879 & 0.524 & -0.144 & 0.01 & -1.87 & 0.533 & -0.0335 & 1.96 & -0.545 & 0.035\tabularnewline
 & $\gamma$ & 0.58 & -0.213 & 0.1 & -0.00613 & 0.824 & -0.256 & 0.0152 & -0.846 & 0.245 & -0.0142\tabularnewline
\hline 
\multirow{3}{*}{\textcolor{red}{$K_{\pi h}$}} & $\alpha$ & 0.862 & -0.0022 & 0.0009 & 0.0013 & 0.442 & -0.015 & -0.0267 & -0.529 & -0.17 & 0.0572\tabularnewline
 & $\beta$ & 0.235 & 0.344 & -0.0925 & 0.0065 & -1.67 & 0.471 & -0.021 & 1.88 & -0.432 & 0.0103\tabularnewline
 & $\gamma$ & 1.02 & -0.252 & 0.128 & -0.0089 & 1.92 & -0.674 & 0.0404 & -2.43 & 0.747 & -0.04\tabularnewline
\hline 
\multirow{3}{*}{\textcolor{red}{$K_{h\pi}$}} & $\alpha$ & 0.862 & -0.0242 & 0.0447 & -0.0071 & 0.665 & -0.417 & 0.024 & -0.613 & 0.123 & 0.0197\tabularnewline
 & $\beta$ & 0.235 & 0.339 & -0.1 & 0.011 & -1.88 & 0.785 & -0.058 & 2.09 & -0.765 & 0.0505\tabularnewline
 & $\gamma$ & 1.02 & -0.156 & 0.104 & -0.0107 & 1.27 & -0.737 & 0.0671 & -1.5 & 0.832 & -0.0731\tabularnewline
\hline 
\multirow{3}{*}{$K_{\pi\pi}$} & $\alpha$ & 1 & 0 & 0 & 0 & 0 & 0 & 0 & 0 & 0 & 0\tabularnewline
 & $\beta$ & 1.18 & 0.0688 & 0.0052 & -0.0001 & 0.107 & -0.0393 & 0.0035 & -0.27 & 0.087 & -0.0063\tabularnewline
 & $\gamma$ & 0.705 & -0.0044 & 0.021 & -0.0011 & 0.0288 & -0.0239 & 0.0005 & -0.0334 & 0.0192 & -0.0003\tabularnewline
\hline 
\end{tabular}\caption{Coefficients in Eq.~\eqref{Ga} for the fitted parameters in Eq.~\eqref{Kred}. }
\label{t:Kred}
\end{table}

The simple fitted kernel \eqref{Kred} with the coefficients \eqref{Ga}
is tested for $AZ^{2}=3,\,4,\,8,\,12,\,16,\,32,\,64,\,256,\,1024,\,4096$
and $T_{\mathrm{i}}/T_{\mathrm{e}}=10^{-4},\,10^{-3},\,10^{-2},\,10^{-1},\,1,\,2,\,3,\,\cdots,\,10$.
The kernel yields accurate closures within 5\% errors for $\hat{h}_{h}$
and $\hat{\pi}_{\pi}$ and within 20\% errors for $\hat{h}_{\pi}$
and $\hat{\pi}_{h}$. The errors greater than 10\% for $\hat{h}_{\pi}$
and $\hat{\pi}_{h}$ occur at small closure values only. Since the
major contribution to closures are from the diagonal elements $\hat{h}_{h}$
and $\hat{\pi}_{\pi}$, the total closures computed from the fitted
kernels \eqref{Kred} are expected to be accurate to within 10\% error
in most cases. 

\section{Summary and future work\label{sec:Dis}}

For ion parallel closures, the heat flow and viscosity closures are
expressed by kernel weighted integrals of temperature and flow velocity
gradients. Simple fitted kernels are obtained by solving the linearized
parallel moment equations for arbitrary atomic weights and charge
numbers. This work together with previous work on electrons \citep{Ji2016KHN}
completes parallel closures for fully ionized electron-ion plasmas
for cases where the magnetic field strength does not vary significantly
along the field line. 

The moment method can be applied to the Landau fluid closures in Ref.~\citep{Joseph2016D}
to obtain the exact linear response for arbitrary collisionality.
In the Landau fluid models, the parallel moments are decomposed into
parallel and perpendicular parts. Therefore the Landau fluid closures
can be constructed as linear combination of parallel moments with
higher order moments included. The Landau fluid closures for the $3_{\|}+1_{\perp}$
model obtained from the moment method will be presented in the near
future.

While the linearized moment equations allow analytical expressions
of the linear response theory, they do not capture coupling effects
between temperature and magnetic field gradients and the moments.
The linear response theory should be a good approximation whenever
the variations in the temperature and magnetic field along a field
line are small. For large variations of temperature and magnetic field,
efforts to include coupling effects of magnetic-field inhomogeneity
and temperature variations are ongoing. 

\section*{Acknowledgments}

The research was supported by the U.S. DOE under grant nos. DE-SC0014033,
DE-SC0016256, DE-FC02-08ER54973, and DE-FG02-04ER54746, and by the
project PE15090 of Korea Polar Research Institute. This work was performed
in conjunction with the Plasma Science and Innovation (PSI) center
and the Center for Extended MHD Modeling (CEMM).

\end{document}